\documentstyle[12pt]{article}
\begin{document}
\vskip 72pt
\centerline{\bf SUPERHEATED DROP NEUTRON SPECTROMETER}
\vskip 36pt
\centerline{Mala Das, B. K. Chatterjee, B. Roy and S.C. ROY}
\centerline{\it Department of Physics, Bose Institute, 
Calcutta 700009, India}
\vskip 36pt
\noindent 

{\bf1.Introduction:}
	
	The Superheated Drop Detector or SDD invented by Apfel in 1979[1] is 
one of the useful detector in neutron detection. The basic principle of 
operation of this detector is same as bubble chamber.  Here the superheated drops are 
suspended in dust free visco-elastic gel medium. Upon nucleation by e
nergetic radiations the drops form bubbles and the drops nucleate independent 
of each other. So one nucleation does not consume the whole liquid and the 
repressurisation process that is needed in bubble chamber, is not required 
here. This is an advantage of SDD over bubble chamber. Each of the drops 
store mechanical energy in it and it is released when triggered by radiations. 
The superheated liquid can be prepared by increasing the temperature of the 
liquid at a given pressure or alternatively it can be prepared by lowering 
the pressure of the liquid at a given temperature. This detector can be made 
on polymer matrix also where the bubbles formed after nucleation of drops are 
tightly bound as was done by Ing and his group, called the "Bubble Detector" 
(BD). Here the nucleation is observed by counting visually the number of 
bubbles trapped in the gel[2]. The test liquid remains in glass tube under 
pressure created by another liquid and just before the experiment, 
the liquid is sensitized by unscrewing  the cap of the tube and allow the 
liquid to become superheated. These SDD/BD serves as an excellent detector 
for the detection of neutrons.	There are different detecting systems by which the nucleation in 
SDD can be measured. One way is to count acoustically the pulses produced by 
drops vaporization with the help of a piezo electric transducer and a drop 
counter[3]. Another way is to measure the volume of the vapor formed upon 
nucleation by a passive method. This system consists of  vertical graduated 
pipette[4] or horizontal glass tubes placed on a graduated platform[5,6] with 
an indicator (gel piston or coloured water column) indicating the volume of 
the vapor formed upon nucleation. The sensitivity of this type of detector 
can be varied by changing the diameter of the glass tube. This system does not 
require any power source and can be used as an alarm dosimeter, in area 
monitoring etc. The third way is to count visually the bubbles trapped in 
hard polymer matrix[2]. The suitability of using superheated drops as a 
neutron dosimeter[7,8,9,10] has already been established. It is a very 
sensitive neutron dosimeter and can measure the neutron dose as low as 
0.1$\mu$Sv. The SDD/BD neutron dosimeters are now commercially available from Apfel 
Enterprises Inc, USA and from Bubble Technology Industries Ltd., Canada.
\vskip 24pt
\noindent 

{\bf2. Principle of neutron spectrometry :}

	Since its discovery, attempts have been made on the  application of 
this detector in neutron spectrometry. There is a minimum energy required for 
nucleation at a given temperature below which no nucleation occurs. This 
minimum energy is called the threshold energy for nucleation. The threshold 
energy decreases as the degree of superheat of the liquid increases. The 
degree of superheat of a liquid is the difference between the vapor pressure 
of the liquid at a given temperature and the ambient pressure or the 
difference between the boiling point of the liquid and the ambient temperature.
Therefore liquid with lower boiling point possess higher degree of superheat 
at a given 
temperature and as the ambient temperature increases the liquid becomes more 
and more superheated. This property of the superheated liquid is being 
utilized to develop the neutron spectrometry. There are different ways by 
which superheated drops can be used in neutron spectrometry. One of the ways 
is to use the different superheated liquids of  different degree of superheat 
and the threshold neutron energies can be obtained by irradiating the 
detectors with different monochromatic neutron sources [11]. Other way is to 
use the same detector operating at different temperature. Since the threshold 
energy depends on the operating temperature of the detector, so by varying 
the temperature of the detector suitably, neutrons of different energies can 
be detected as was done by d'Errico et. al.[12]. They have used two superheated 
liquids operating at four different temperatures to obtain eight different 
threshold neutron energies . It is to be noted that in order to get a good 
resolution of the spectrum, temperature variation at a close grid is necessary.

\vskip 24pt
\noindent
{\bf3. Present work :}
	Now there is a different approach by which the neutron energy spectrum 
can be obtained from the temperature dependence of threshold energy of SDD. 
After the interactions of the neutrons with the nuclei of the constituting 
atoms of superheated liquid, ions of different energies are formed. The ion 
having the highest value of LET (dE/dx) in the liquid, will play the major 
role in nucleation. Another important point is that there is a specific 
length L, along the ions track, the energy ($E_c$) deposited over that length 
will contribute significant role in nucleation. Actually a very small fraction 
of the deposited energy is usually used up in nucleation i.e. W/$E_c$ is very 
small and this ratio is called the thermodynamic efficiency of nucleation ((T). 
After the deposition of energy by the ions, nucleation occurs with the 
formation of a critical size vapor bubble of radius rc inside the liquid drop. 
It is suggested that L  = 2 $r_c$ [13,14] and $E_c$ can be expressed as 
$E_c$ = 2 $r_c$ dE/dx Therefore,W = ${{\eta}_T {E_c}}$  or	 W = 2 ${{\eta}_T {r_c}}$ dE/dx              

or        W/$r_c$ = k dE/dx,            where k = 2 ${\eta}_T$W and ${r_c}$ both are functions of temperature and dE/dx is a function of the 
energy of the projectile ions in superheated liquid, which can be converted 
to the energy of the incident neutrons. So this equation relates the threshold 
neutron energy for nucleation to the ambient temperature. This enables one to 
convert the temperature of the detector to the energy of the incident neutrons. 
Therefore using the above equation as a working equation, neutron energy spectrum 
can be obtained by observing the SDD response at different temperature. This 
gives an important application of SDD in neutron spectrometry. The nucleation 
rate in superheated drops is proportional to the total volume of the drops (V), 
incident neutron flux $\psi$, neutron-nucleus interaction cross section $\sigma$ 
and to the neutron detection efficiency $\eta$ of SDD. Neutron detection 
efficiency $\eta$, 
is defined as the fraction of the incident neutrons which are used up in 
nucleation. By observing the nucleation rate in superheated drops, 
$\eta$ can be obtained from the known values of V, $\psi$ and $\sigma$. If one measures 
$\eta$ at different temperatures, the derivative of $\eta$ against temperature 
resembles the neutron energy spectrum of the source. The temperature axis can be converted 
to the neutron energy following the method discussed here. For a given neutron 
energy spectrum, at low temperature only the high energy neutrons take part 
in nucleation and as temperature increases, threshold energy decreases and so 
in addition to the high energy neutrons, low energy neutrons are also detected. 
So for a polychromatic source, $\eta$ should increase with temperature. When all 
the neutrons in the spectrum contribute in nucleation, $\eta$ should be constant 
with temperature because no more neutrons are left to be detected. For a 
monochromatic source, there is only one sharp increase of $\eta$ at a particular 
temperature corresponding to the energy and it 
should be constant for the other temperatures. This detector can be made 
sensitive to different ranges of neutron energies as the user's choice by 
varying the temperature of the liquid. This detector can detect neutrons with 
energies ranging from thermal to fast energy. We have tested the present 
principle of spectrometry with a 3 Ci Am-Be neutron source[15]. The temperature 
was varied from -17$^oC$ to about 60$^oC$ with the help of an indigenously made 
temperature controller. There is a fair agreement between the neutron energy 
spectrum of Am-Be obtained from our experiment and the available spectrum of 
the source. 	The main advantage of this type of spectrometer is that it is 
easy to prepare, low cost of the materials and no need of power supply. 
Nowadays, Superheated Drop Detectors are widely used in the determination of 
neutron spectra in space, at high altitude studies, gamma detection, detection 
of radon, for cold dark matter search, charged particle detection etc. Besides 
these applications, the radiation induced nucleation in superheated drop 
detector is itself a very interesting field of research.
\vskip 24pt
\noindent

{\bf References :-}\noindent{1. R. E. Apfel (1979) U. S. Patent 4,143,274.}\\\noindent{2. H. Ing and H. C. Birnboim (1984) Nucl. Tracks and Rad. Meas. 8, 285.}  \noindent{3. R. E. Apfel and S. C. Roy (1983) Rev. Sci. Instrum. 54, 1397.}\\ \noindent{4. R. E. Apfel (1992) Rad. Prot. Dos. 44, 343.}\\\noindent{5. B. Roy, B. K. Chatterjee, Mala Das and S. C. Roy (1998)  Rad. Phys. Chem. 51, 473.}\\ \noindent{6. Mala Das, B. Roy, B. K. Chatterjee and S. C. Roy (1999) Rad. Meas. 30, 35.}\\\noindent{7. R. E. Apfel and S. C. Roy (1984) Nucl. Inst. Meth. 219, 582.}\\\noindent{8. R. E. Apfel and Y. C. Lo (1989) Health Phys. 56, 79.}\\\noindent{9. S. C. Roy, R. E. Apfel and Y. C. Lo (1987) Nucl. Inst. Meth. A255, 199.}\\\noindent{10. H. Ing (1986) Nuclear Tracks 12, 49.}\\\noindent{11. H. Ing, R. A. Noulty and T. D. Mclean (1995) Rad. Meas. 27, 1.}\\\noindent{12. F. d'Errico Rad. Prot. Dos. (1995) Rad. Prot. Dos. 61, 159.}\\ \noindent{13. R. E. Apfel, S. C. Roy and Y. C. Lo (1985) Phys. Rev. A 31, 3194.}\\ \noindent{14. M. J. Harper and M. E. Nelson (1990) Rad. Prot. Dos. 47, 535.}\\ \noindent{15. Mala Das, B. K. Chatterjee, B. Roy and S. C. Roy (1999) Nucl. Inst. Meth A 
(submitted).}\\
\end {document}